\documentclass{article}

\usepackage{graphicx}  
\usepackage{dcolumn}   
\usepackage{bm}        
\usepackage{amssymb}   

\usepackage{amsmath,amsfonts,
latexsym,
cite,bm,
} 
\usepackage{slashed} 
\usepackage{relsize} 
\usepackage{prodint} 

\usepackage{hyperref} 

\usepackage{color}

\usepackage{soul} 

\author{Igor Kanatchikov$^{1,2}$
\\
$^{1}$ 
University of St Andrews, St Andrews KY16 9SS, UK \\
$^{2}$  
National Quantum Information Center in Gda\'nsk (KCIK), \\ 
81-831 Sopot, Poland
}

\date{}

\title{\bf On the precanonical structure of the Schr\"odinger 
wave functional in curved space-time }

\begin{document}
\maketitle
\begin{abstract} 
The functional Schr\"odinger equation in curved space-time is derived from the 
manifestly covariant precanonical Schr\"odinger equation. The Schr\"odinger wave functional is expressed as the trace of the multidimensional product integral of precanonical wave function restricted to a field configuration. The functional Schr\"odinger representation of QFT in curved space-time appears as a singular limiting case of a formulation based on precanonical quantization, which leads to a hypercomplex generalization of quantum formalism in field theory.  
\vspace{6pt} 

\vspace{2pt}

 
\end{abstract}


\newcommand{\beq}{\begin{equation}}
\newcommand{\eeq}{\end{equation}}
\newcommand{\beqa}{\begin{eqnarray}}
\newcommand{\eeqa}{\end{eqnarray}}
\newcommand{\nn}{\nonumber}
\newcommand{\bew}{\begin{widetext}}
\newcommand{\eew}{\end{widetext}}

\newcommand{\half}{\frac{1}{2}}

\newcommand{\xt}{\tilde{X}}

\newcommand{\uind}[2]{^{#1_1 \, ... \, #1_{#2}} }
\newcommand{\lind}[2]{_{#1_1 \, ... \, #1_{#2}} }
\newcommand{\com}[2]{[#1,#2]_{-}} 
\newcommand{\acom}[2]{[#1,#2]_{+}} 
\newcommand{\compm}[2]{[#1,#2]_{\pm}}

\newcommand{\lie}[1]{\pounds_{#1}}
\newcommand{\co}{\circ}
\newcommand{\sgn}[1]{(-1)^{#1}}
\newcommand{\lbr}[2]{ [ \hspace*{-1.5pt} [ #1 , #2 ] \hspace*{-1.5pt} ] }
\newcommand{\lbrpm}[2]{ [ \hspace*{-1.5pt} [ #1 , #2 ] \hspace*{-1.5pt}
 ]_{\pm} }
\newcommand{\lbrp}[2]{ [ \hspace*{-1.5pt} [ #1 , #2 ] \hspace*{-1.5pt} ]_+ }
\newcommand{\lbrm}[2]{ [ \hspace*{-1.5pt} [ #1 , #2 ] \hspace*{-1.5pt} ]_- }

\newcommand{\pbr}[2]{ \{ \hspace*{-2.2pt} [ #1 , #2\hspace*{1.4 pt} ] 
\hspace*{-2.3pt} \} }
\newcommand{\nbr}[2]{ [ \hspace*{-1.5pt} [ #1 , #2 \hspace*{0.pt} ] 
\hspace*{-1.3pt} ] }

\newcommand{\we}{\wedge}
\newcommand{\nbrpq}[2]{\nbr{\xxi{#1}{1}}{\xxi{#2}{2}}}
\newcommand{\lieni}[2]{$\pounds$${}_{\stackrel{#1}{X}_{#2}}$  }

\newcommand{\rbox}[2]{\raisebox{#1}{#2}}
\newcommand{\xx}[1]{\raisebox{1pt}{$\stackrel{#1}{X}$}}
\newcommand{\xxi}[2]{\raisebox{1pt}{$\stackrel{#1}{X}$$_{#2}$}}
\newcommand{\ff}[1]{\raisebox{1pt}{$\stackrel{#1}{F}$}}
\newcommand{\dd}[1]{\raisebox{1pt}{$\stackrel{#1}{D}$}}
\newcommand{\der}{\partial}
\newcommand{\oo}{$\Omega$}
\newcommand{\Om}{\Omega}
\newcommand{\om}{\omega}
\newcommand{\eps}{\epsilon}
\newcommand{\si}{\sigma}
\newcommand{\Lm}{\bigwedge^*}

\newcommand{\inn}{\hspace*{2pt}\raisebox{-1pt}{\rule{6pt}{.3pt}\hspace*
{0pt}\rule{.3pt}{8pt}\hspace*{3pt}}}
\newcommand{\sro}{Schr\"{o}dinger\ }
\newcommand{\vol}{\omega}
               \newcommand{\dvol}[1]{\der_{#1}\inn \vol}

\newcommand{\bd}{\mbox{\bf d}}
\newcommand{\bder}{\mbox{\bm $\der$}}
\newcommand{\bI}{\mbox{\bm $I$}}

\newcommand{\gammat}{\; \widetilde{\gamma}{}}
\newcommand{\be}{\beta} 
\newcommand{\ga}{\gamma} 
\newcommand{\de}{\delta} 
\newcommand{\Ga}{\Gamma} 
\newcommand{\gmu}{\gamma^\mu}
\newcommand{\gnu}{\gamma^\nu}
 \newcommand{\ka}{\varkappa} 
 \newcommand{\la}{\lambda}
\newcommand{\hka}{\hbar \kappa}
\newcommand{\al}{\alpha}
\newcommand{\lapl}{\bigtriangleup}
\newcommand{\psib}{\overline{\psi}}
\newcommand{\Psib}{\overline{\Psi}}
\newcommand{\Phib}{\overline{\Phi}}
\newcommand{\derts}{\stackrel{\leftrightarrow}{\der}}
\newcommand{\what}[1]{\widehat{#1}}

\newcommand{\gammab}{\underline{\gamma}}

\newcommand{\bx}{{\bf x}}
\newcommand{\bk}{{\bf k}}
\newcommand{\bq}{{\bf q}}

\newcommand{\omk}{\omega_{\bf k}} 
\newcommand{\lpl}{\ell}
\newcommand{\zb}{\overline{z}} 

\newcommand{\deltab}{\bar \delta}

\newcommand{\dv}{\mbox{\sf d}}

\newcommand{\SE}{\mbox{Schr\"odinger equation\ }}

\newcommand{\deltt}{\bm{\delta}}   

 \newcommand{\BPsi}{\text{\boldmath$\Psi$}} 
 \newcommand{\BPhi}{\text{\boldmath$\Phi$}} 
 \newcommand{\BXi}{\text{\boldmath$\Xi$}} 

\newcommand{\BH}{{\bf H}} 
\newcommand{\BS}{{\bf S}} 
\newcommand{\BN}{{\bf N}} 

\newcommand{\rd}{\mathrm{d}}
\newcommand{\ri}{\mathrm{i}}
\newcommand{\Tr}{\mathrm{Tr}} 

\newcommand{\boldx}{{\bx}} 
\newcommand{\equn}{{\rm equation}}
\newcommand{\ota}{{\rm obtain}}
\newcommand{\dwh}{{\rm DW Hamiltonian}}
\newcommand{\fd}{{\rm field}}
\newcommand{\con}{{\rm connection}}
\newcommand{\bewn}{{\rm between}}
 \newcommand{\vsh}{{\rm vanish}}
 \newcommand{\lmt}{{\rm limit}}
 \newcommand{\tmm}{{\rm term}}
 \newcommand{\cse}{{\rm case}}
 \newcommand{\Sch}{{\rm Schr\"{o}dinger}}
 \newcommand{\wnn}{{\rm when}}
 \newcommand{\ltr}{{\rm latter}}
 \newcommand{\rsd}{{\rm restricted}}
  \newcommand{\drv}{{\rm derivat}}
 \newcommand{\clf}{{\rm Clifford}}
\newcommand{\fst}{{\rm first}}
\newcommand{\rln}{{\rm relation}}
\newcommand{\dfrt}{{\rm different}}
\newcommand{\rsp}{{\rm respect}}
\newcommand{\by}{{\rm by}}
\newcommand{\spt}{{\rm space-time}}
\newcommand{\crsp}{{\rm correspond}}
\newcommand{\trt}{{\rm treat}}
\newcommand{\hlt}{{\rm Hamilton}} 
\newcommand{\and}{{\rm and}}
\newcommand{\qnt}{{\rm quant}} 
\newcommand{\myy}{{\rm we}}
\newcommand{\imls}{{\rm implies}}
\newcommand{\dnt}{{\rm denote}}
\newcommand{\gma}{\gamma}
\newcommand{\Si}{\Sigma}


\section{\large Introduction}  

 Precanonical quantization 
 \cite{berlin,bl,lz,gequ,my-ehr} 
 is the approach to field quantization based on the De Donder-Weyl (DW) generalization of Hamiltonian formalism to field theory  \cite{kastrup} which does not require the space+time decomposition and treats all space-time variables on equal footing.  
Despite the DW theory has been known since the 1930-es and it was considered as a possible basis of field quantization by Hermann Weyl himself \cite{weyl34}, 
its various mathematical structures have been studied starting from the late 1960-es (with the relevant notion of the multisymplectic structure coined in Poland \cite{kij}), it is the 
structure of the Poisson-Gerstenhaber algebra of Poisson brackets 
defined on differential forms found within the DW Hamiltonian 
formulation in 
 \cite{gequ,pg1,pg2} 
which has proven to be suitable for a new approach 
to field quantization. Further discussion of this bracket 
or its different treatments and generalizations can be found e.g. 
in 
 \cite{jo1,jo2,loday,rmr,dkp00,dirbr}. 
 It also has been instrumental 
in recent discussions of various classical field theoretic models of gravity and gauge theories in 
 \cite{molgado1,molgado2,molgado3}. 
 %

Applications of precanonical quantization so far include quantum gravity 
in metric \cite{pqgr,pqgr4,pqgr3,pqgr2}  
and vielbein variables 
 \cite{qg,qg2,qg3,qg4,mg-ehr},  
 and quantum Yang-Mills theory 
 \cite{my-ym1,ym-mass,romp2018}. 
However, the connection with the standard techniques 
and concepts of QFT still remains not sufficiently explored. 

Many aspects of the relations between the DW Hamiltonian theory 
and the canonical Hamiltonian formalism have been studies since 
the early 1970-es including 
 \cite{kij,sniat,gotay,pg1,jo1,jo2,my-pla,riahi}. 
The nature of those relations is that, typically, a 
covariant geometrical object from the  DW Hamiltonian theory 
leads to its canonical counterpart after the space+time decomposition, restriction to a subspace representing the Cauchy data and then integration over it. Hence the name ``precanonical" for the DW Hamiltonian formalism and the related quantization. 
%
%

On the quantum level, the connection between precanonical quantization and the functional Schr\"o\-dinger representation 
of QFT \cite{hatf} was found for scalar fields 
 \cite{my-pla,atmp1,atmp2} 
 and YM fields 
 \cite{my-ym1,ym-mass,romp2018} 
in flat space-time. As a step towards understanding 
the connections between precanonical quantization of gravity   
 \cite{pqgr,pqgr4,pqgr3,pqgr2,qg,qg2,qg3,qg4} 
and the existing approaches based on canonical quantization \cite{rovelli,kiefer} 
we have also explored a possible extension of those results to curved space-time 
 \cite{static18,nonstatic18,nonstatic19}. 
 The present discussion is a concise presentation of 
 those papers. 

As in the case of flat space-time, it will be shown that the functional Schr\"odinger representation in curved space-times \cite{frs,shor1} emerges from the description derived from precanonical quantization as a singular limiting case.


\newcommand{\dmu}{\frac{d}{d x^\mu}}

\section{\large Precanonical description of quantum scalar theory}

Let us start from the Lagrangian density 
${\frak L} = \frac{1}{2} \sqrt{g}g^{\mu\nu} \der_\mu \phi \der_\nu \phi - \sqrt{g} V(\phi)$,  
where $g_{\mu\nu}(x)$ is the space-time metric, 
$g: = \vert \det (g_{\mu\nu}) \vert$.
It defines the densities of polymomenta 
$\frak{p}^\mu := \frac{\der \frak{L}}{\der \der_\mu \phi} = \sqrt{g}g^{\mu\nu} \der_\mu \phi$  
and the DW Hamiltonian density 
${\frak H} = \sqrt{g} H 
:=\frak{p}^\mu \der_\mu \phi (\frak{p}) - {\frak L} 
= \frac{1}{2\sqrt{g}} g_{\mu\nu} \frak{p}^\mu \frak{p}^{\nu} + \sqrt{g} V(\phi)$.
Then the field equations take the DW Hamiltonian form
\beq \label{dw-curved}
d \frak{p}^\mu (x)/dx^\mu = - {\der \frak{H}}/{\der \phi}, \qquad 
d \phi(x)/ dx^\mu  =  {\der \frak{H}} / {\der \frak{p}^\mu}.   
\eeq 

The Poisson bracket operation defined by the weight $+1$ density valued $x$-dependent polysymplectic structure 
 $\Omega  = \rd \frak{p}^\mu\we \rd\phi\we \varpi_\mu$, where 
 $\varpi_\mu := \der_\mu \inn\ (dx^0\we dx^1\we ... \we dx^{n-1})$, 
yields \cite{my-ehr,pg2,pg1} 
\beq
\pbr{\frak{p}^\mu \varpi_\mu}{\phi } = 1, \quad 
\pbr{\frak{p}{}^\mu\varpi_\mu}{\phi\varpi_\nu } = \varpi_\nu, 
\quad 
\pbr{\frak{p}^\mu}{\phi \varpi_\nu } = \delta^\mu_\nu .
\eeq
These fundamental Poisson brackets are quantized 
according to the modified Dirac quantization rule  
\beq 
[\hat{A}, \hat{B}] = - \ri\hbar\sqrt{g} \what{\pbr{A}{B}} , 
\eeq
that leads to the following 
representations of precanonical quantum operators  
\beq
\hat{\frak{p}}{}^\mu  = -\ri\hbar\ka\sqrt{g} \gamma^\mu \der_\phi , 
\quad 
\hat{\varpi}_\nu = \frac1\ka \gamma_{\mu} , 
\quad 
\what{H} = -\frac{1}{2}\hbar^2\ka^2\der_{\phi\phi} + V(\phi) , 
\eeq
where $\gamma^\mu$ are $x$-dependent Dirac matrices,  
$\gamma^\mu\gamma^\mu+ \gamma^\mu\gamma^\mu = 2g^{\mu\nu}$, 
      the composition of operators is the symmetrized 
      Clifford (matrix) product and $\ka$ is a ultraviolet scale 
      appearing on dimensional grounds. 

The curved space-time version of 
 the precanonical Schr\"odinger equation 
takes the form 
\beq \label{crv-ns}  
\ri\hbar\ka\gamma^\mu \nabla_\mu\Psi = \what{H}\Psi ,  
\eeq
where $\nabla_\mu := \der_\mu + \omega_\mu (x) $ is a covariant derivative of Clifford algebra valued wave functions $\Psi (\phi, x^\mu)$
 with the spin connection matrix $\omega_\mu = \frac14 \omega_\mu^{AB} 
 \underline{\gamma}{}_{AB}$ acting on Clifford-valued 
    $\Psi$ via the commutator product, where 
    $\underline{\gamma}{}_{A}$ are the Minkowski space Dirac matrices.



\newcommand{\oldatxta}{
The  precanonical quantization of a scalar field $\phi(x)$ in curved space-time with the metric $g_{\mu\nu} (x)$   (cf.~\cite{my-ehr,mg-ehr}) leads to the 
description  in terms of a  Clifford-algebra-valued wave function $\Psi(\phi, x^\mu)$ 
on the finite-dimensional bundle over space-time 
with the local coordinates $(\phi,x^\mu)$.  
The precanonical wave function $\Psi$ satisfies 
the partial derivative precanonical Schr\"odinger equation (pSE)
\beq  \label{crv-ns}
\ri\hbar \gamma^\mu (x) 
 \nabla_\mu \Psi = 
 \left(- \frac{1}{2} \hbar^2\varkappa \frac{\der^2}{\der \phi^2 } 
+ \frac1\ka V(\phi)  \right)\Psi =: \frac1\ka\what{H}\Psi \,, 
\eeq
where 
$\gamma^\mu (x)$ are the  curved space-time Dirac matrices such that 
$\gamma^\mu (x)\gamma^\nu (x) + \gamma^\nu (x) \gamma^\mu (x) = 2g^{\mu\nu}(x)$, 
\beq
\nabla_\mu := \der_\mu + \omega_\mu(x)
\eeq 
is the covariant derivative  with the spin-connection matrices 
$\omega_\mu (x) = \frac14 \omega_\mu{}_{AB}(x)\underline{\gamma}{}^{AB}$ 
(\cite{pol}) 
acting on Clifford-algebra-valued wave functions 
 by the commutator product, 
 and 
$\underline{\gamma}{}^{A}$ are the constant Dirac matrices 
$\underline{\gamma}{}^{A} \underline{\gamma}{}^{B} 
+ \underline{\gamma}{}^{B} \underline{\gamma}{}^{A} = 2 \eta^{AB}$ 
related to the Minkowski metric $\eta^{AB}$.

Note that plain capital Greek letters like $\Psi$ and $\Phi$ 
denote wave functions on a finite 
dimensional space of $\phi$ and $x^\mu$ 
and the boldface capital Greek letters like $\BPsi$ and $\BPhi$ denote functionals of field configurations $\phi(\bx)$. We also set $\hbar=1$ from now on.  
   
The operator $\what{H}$ 
in~(\ref{crv-ns}) is the 
De Donder-Weyl (DW) Hamiltonian operator constructed according to the procedure of 
precanonical quantization~\cite{bl,lz,my-ehr,mg-ehr}. In the expression of $\what{H}$ there appears an ultraviolet parameter $\ka$ of the dimension of the inverse spatial volume.  This parameter typically appears in the representations of precanonical 
quantum operators ~\cite{berlin,bl,lz,my-ehr}. 
The DW Hamiltonian operator $\what{H}$ for scalar fields in curved space-time 
coincides with its flat space-time analogue (cf.~\cite{berlin,bl,lz,my-ehr}),  
hence the curved space-time  manifests itself only throught the curved space-time 
Dirac operator in the left hand side of (\ref{crv-ns}).   

 }

\section{\large Relating the functional Schr\"odinger picture and precanonical quantization} 

The issue we would like to address here is a relation between 
the description of quantum fields derived from precanonical quantization and the standard QFT based on canonical quantization. 
More specifically, we would like to generalize the relation found 
in flat space-time in the case of quantum scalar field theory 
\cite{my-pla,atmp1,atmp2} and quantum YM theory \cite{my-ym1,romp2018} 
to curved space-time. Our presentation here follows \cite{static18,nonstatic18,nonstatic19}. 

In curved space-time, quantum scalar field can be described 
in terms  of the  wave functional $\BPsi([\phi(\bx)],t)$ of 
field configurations $\phi(\bx)$ at the time t which obeys the 
 Schr\"odinger equation~\cite{frs,shor1} 
\beq \label{fs}
\ri\hbar \der_t \BPsi = 
 \int \! d\bx\, \sqrt{g}
 \left (  \frac{\hbar^2}{2}\ \frac{g_{00}}{g}\frac{\delta^2}{\delta \phi(\bx)^2} 
- \frac{1}{2} g{}^{ij} \der_i\phi(\bx)\der_j\phi(\bx) + V(\phi)
 \right) \BPsi , 
\eeq 
where the space-time coordinates adapted to 
the codimension-one space-like foliation are used such as 
$g_{0i}=0$ and $g_{ij}$ is 
the induced metric on the space-like leaves of the foliation, and 
 $x^\mu = (t,\bx)= (t, x^i)$. As in all our poapers, 
 bold capital Greek letters like 
 $\BPsi, \BPhi$ denote functionals and plain capital Greek letters like 
 $\Psi, \Phi$ denote (precanonical) wave functions. 

The problem of relating the  canonical Schr\"odinger equation in functional derivatives with the partial derivative precanonical \SE\ (\ref{crv-ns}) 
manifests itself already on the classical level:  
in \cite{my-pla} we have shown how the partial derivative DW Hamilton-Jacobi equation is related to the canonical Hamilton-Jacobi equation in variational 
derivatives. Our initial consideration in flat space-time 
recently was extended to curved space-time and general relativity 
in \cite{riahi}. 

On the quantum level, the idea is that the wave functional is a functional of field configurations because it is a composed functional of precanonical wave function $\Psi(\phi^a,x)$ 
restricted to the 
section $\Sigma$ in the total space of the 
bundle with the coordinates $(\phi^a,x)$ which 
represents a field configuration 
$\phi(\bx)$ at the moment of time $t$: 
$\Psi_\Sigma := \Psi (\phi^a = \phi^a(\bx), \bx, t)$, i.e. 
\beq \label{psib}
\BPsi([\phi(\bx)],t) = \BPsi ([\Psi_\Sigma (\bx,t), \phi (\bx)]) .
\eeq 
In this case the time dependence of the wave functional $\BPsi$ 
originates from the time dependence of precanonical wave function restricted to $\Sigma$ and the variational derivatives of $\BPsi$ 
can be related to the partial derivatives of $\Psi_\Sigma$. 
By denoting  $\BPhi^{}({\bx})
:= \frac{\delta \BPsi}{\delta\Psi^T_\Sigma (\bx)} $ 
($\Psi^T$ is the transpose of $\Psi$),  we obtain 
\begin{align} 
\label{dtps}
 \ri\der_t \BPsi &= 
 {\Tr} \int\! d\bx\, 
 \left \{ 
 \BPhi^{}({\bx})
\, \ri\der_t \Psi_\Sigma (\bx, t)  
\right \} , \\
\label{dbps}
 \frac{\delta \BPsi }{\delta \phi(\bx)}
&=
\Tr\left \{
 \BPhi^{}({\bx})
\der_\phi \Psi_\Sigma (\bx)
\right \}
+  \frac{\deltab \BPsi }{\deltab \phi(\bx)^{{}^{}}}, \\[1ex]
\begin{split}
\label{dbps2} 
\frac{\delta^2 \BPsi }{\delta \phi(\bx)^2}
&=
\Tr\left \{
 \delta(\mathbf{0}) \BPhi^{}({\bx}) \der_{\phi\phi} \Psi_\Sigma (\bx) 
+ 2 \frac{ \deltab \BPhi(\bx)}{\deltab \phi(\bx)}   ~\der_\phi \Psi_\Sigma (\bx) \right \}
  \\[1ex]
 &\quad+\Tr \, \Tr \left \{
 \frac{\delta^2 \BPsi}{\delta \Psi^T_\Sigma(\bx)\otimes\delta\Psi^T_\Sigma(\bx)}
~\der_\phi \Psi_\Sigma (\bx)
\otimes  \der_\phi \Psi_\Sigma  (\bx) 
\right \}  
+  \frac{\deltab^2 \BPsi }{\deltab \phi(\bx)^2}
 , 
\end{split}
\end{align}
where 
 $\der_\phi \Psi_\Sigma(\bx) := ({\der}\Psi  /{\der \phi}) |_\Sigma (\bx) , \quad 
\der_{\phi\phi} \Psi_\Sigma (\bx) := ({\der^2}\Psi /{\der \phi^2})  |_\Sigma (\bx)$, 
$\deltab$ is the { partial} functional derivative with respect to $\phi(\bx)$, 
 as opposite to the { total} functional derivative $\delta$,  
and $\delta(\mathbf{0})$ is a regularized value of 
$\delta \Psi_\Sigma (\bx)/ \delta \Psi^T_\Sigma (\bx')$ at $\bx = \bx'$, 
which can be defined using a point splitting or lattice regularization 
to make sense of $(n-1)$-dimensional delta-function $\delta(\bx-\bx')$ at equal 
points. 

The time derivative 
of $\Psi_\Sigma$ is determined by the restriction of 
precanonical Schr\"odinger equation 
(\ref{crv-ns}) in  space+time decomposed form 
to  $\Sigma$:  
\beq \label{eq6n}
\ri\der_t \Psi_\Sigma = -\ri\gamma_0\gamma^i \left ( \frac{d}{dx^i} 
- \der_{i} \phi (\bx) \frac{\der}{\der \phi } \right ) \Psi_\Sigma 
- \ri \gamma_0 \gamma^{i} [\omega_i{}, \Psi_\Sigma ]  
- \ri  [\omega_0{}, \Psi_\Sigma ]
+ \frac{\gamma_0}{\ka} \what{H} 
\Psi_\Sigma, 
\eeq 
where  $\frac{d}{dx^i} := \der_i + \der_i \phi(\bx)  \frac{\der}{\der \phi} 
+ \der_i \phi_{k}(\bx) \frac{\der}{\der \phi_{,k} } +... $ 
is the total derivative along $\Sigma$ 
with $\phi_{k}$ denoting the fiber coordinates of the 
first-jet bundle of the bundle of field variables 
$\phi$ over space-time such that,  when restricted to 
$\Sigma$, $\phi_{k} = \der_k \phi(\bx)$. 
 
 
By substituting (\ref{eq6n}) to (\ref{dtps}) we obtain 
\begin{align} \label{dt}
\begin{split}
i\der_t \BPsi =
    \Tr \int \!\rd\bx\ 
 \bigg\{
 \BPhi (\bx,t)
\Big( &
\underbrace{-\ri\gamma_0\gamma^i \frac{d}{dx^i} \Psi_\Sigma  (\bx)}_{I} 
 + \underbrace{\ri \gamma_0\gamma^i\der_i \phi(\bx){\der_\phi} \Psi_\Sigma  (\bx)
 \vphantom{\frac{C}{D}}}_{II}
   \\
 \underbrace{\strut - {\ri} \gamma_0 \gamma^{i} [\omega_i{}, \Psi_\Sigma (\bx)]
 \vphantom{\frac{A}{B}} }_{III}  &
 \underbrace{\strut - {\ri}  [\omega_0{}, \Psi_\Sigma ] 
 \vphantom{\frac{A}{B}}
 }_{IV} 
\underbrace{- \gamma_0 \frac{\varkappa}{2} \der_{\phi\phi}\Psi_\Sigma  (\bx)}_{V} 
+ 
  \underbrace{ \gamma_0 \frac{1}{\varkappa} V(\phi(\bx)) \Psi_\Sigma  (\bx)}_{VI}
\Big ) \bigg\} . 
\end{split}
\end{align} 
 
By comparing the term $V$ with the first term in (\ref{dbps2}) and (\ref{fs}) 
we conclude that the relation between them is only possible 
under the limiting mapping $\mapsto$ such that 
\beq  \label{cond}
\gamma_0 \ka  \mapsto \frac{g_{00}}{\sqrt{g}}\delta(\mathbf{0}).
\eeq
We will see in what follows that the same condition also 
appears in order places when we are trying to 
establish a correspondence between the terms in (\ref{dt}) and 
the ones in (\ref{fs}).

The potential term $VI$ in   (\ref{dt}) should reproduce 
the potential term in (\ref{fs}). It is easy to see it is possible 
only if  at any spatial point $\bx$ there is a mapping $\mapsto$ 
such that 
\beq \label{vee}
\Tr \left \{    
\BPhi(\bx) 
 ~ \frac{1}{\ka} \gamma_0 
 \Psi_\Sigma (\bx) \right \}
  \mapsto \sqrt{g}\ \BPsi .
\eeq
The study of this conditions shows~\cite{static18}  
that it can be fulfilled if 
the third term in (\ref{dbps2}) vanishes
and the  limiting condition (\ref{cond}) 
is satisfied.  

Using the fact that $\sqrt{g} = \sqrt{g_{00} h}$, 
where $h:=\vert \det (g_{ij})\vert$, 
and $ \sqrt{g_{00}} \gamma^0 = \underline{\gamma}{}_0$, 
the condition (\ref{cond}) takes the form 
\beq \label{limh}
\underline{\gamma}{}_0\ka \mapsto \delta(\mathbf{0})/\sqrt{h} = \delta^{\mathrm{inv}}(\mathbf{0}) ,
\eeq 
where $\delta^{\mathrm{inv}}(\bx)$ is the invariant $(n-1)$-dimensional delta-function  such that 
$\int\! d\bx \sqrt{h} \delta^{\mathrm{inv}}(\mathbf{x}) =1$.  
 Besides, the above definition of $\delta^{\mathrm{inv}}(\mathbf{\mathbf{x}})$ 
implies that $\delta^{\mathrm{inv}}(\mathbf{\mathbf{0}})$  is the inverse of the invariant volume element $ \sqrt{h} d\bx $. 
Then (\ref{limh}) is equivalent to  
\beq \label{limdx}
\frac{1}{\ka} \mapsto \underline{\gamma}{}_0 \sqrt{h} d\bx . 
\eeq
This interpretation will be used when writing the product integral expressions of the wave functional in terms of precanonical wave function.  

Our next observation is that the second term in (\ref{dbps2}) 
is similar to  the tern $II$ in (\ref{dt}) in that both 
contain $\der_\phi \Psi_\Sigma$ and do not have a counterpart 
in (\ref{fs}). Hence they have to cancel each other at least in 
the limit (\ref{cond}), i.e. 
\beq \label{dypsi} 
 \ri \BPhi^{}({\bx})
 \gamma_0 \gamma^i\der_i\phi(\bx) 
 +  \frac{g_{00}}{\sqrt{g}} 
\frac{\deltab \BPhi (\bx)}{\deltab \phi(\bx)} \mapsto 0, 
\eeq
The solution of (\ref{dypsi}) can be written in the form 
\beq \label{bphi}
\BPhi^{}(\bx) =  \BXi([\Psi_\Sigma];\check{\bx})
e^{-\ri \phi(\bx)\gamma^i\der_i\phi(\bx)/\ka},
\eeq
where  the ``integration constant" $\BXi([\Psi_\Sigma];\check{\bx})$ 
obeys 
$
 \frac{\deltab \BXi([\Psi_\Sigma];\check{\bx})}{\deltab \phi(\bx)}\equiv 0 
$.   
Therefore the required cancellation of the terms with $\der_\phi \Psi_\Sigma(\bx)$ in the limit~(\ref{cond})) fixes the form of the functional $\BPhi(\bx)$.  
 This allows us to express the wave functional $\BPsi$  
 in the form 
\beq \label{bpsi3}
\BPsi \sim  
 \Tr
\left \{\BXi([\Psi_\Sigma];\check{\bx})~
e^{-\ri \phi(\bx)\gamma^i\der_i\phi(\bx)/\ka}~
 \frac{\gamma_0}{\sqrt{g} \ka}
 \Psi_\Sigma (\bx) \right \}_{\mbox{\Large $\rvert$} \scriptstyle
 \ka \stackrel{\mbox{\tiny $$}}{\mbox{\scriptsize $\longmapsto$}} 
 \gamma_0\delta(\mathbf{0})/ \sqrt{g}  },
\eeq  
which is valid at any point  $\bx$; $\sim$ here and in what follows 
is the equality up to a normalization factor  which 
may depend on  $\ka$ and $\sqrt{g}$. 
The notation $\{ ... \}_{\mbox{\large $\rvert$} \scriptstyle
 \ka \stackrel{\mbox{\tiny $$}}{\mbox{\scriptsize $\longmapsto$}} 
 \gamma_0\delta(\mathbf{0}) / \sqrt{g}}$
 means that every appearance of $\ka$ in the expression inside braces 
 is replaced  by  
 $\gamma_0\delta(\mathbf{0}) / \sqrt{g}$ 
 according to the limiting map~(\ref{cond}).

Now, using (\ref{bpsi3}), for  the last term 
in~(\ref{dbps2}) 
we get 
\beq \label{e32}
\frac12 \frac{g_{00}}{\sqrt{g}} \frac{\deltab^2 \BPsi }{\deltab \phi(\bx)^2}
\mapsto
 -  \frac{1}{2}\sqrt{g} g^{ij}\der_i \phi(\bx) \der_j \phi(\bx) \BPsi ,   
\eeq
that reproduces the second term in the functional derivative 
Schr\"o\-dinger equation~(\ref{fs}).

Thus, in the limiting case (\ref{cond}),   
we have derived all terms in (\ref{fs}) 
from (\ref{dt}). However, the terms $I+III+IV$ in 
(\ref{dt}) still have played no role. 
One can argue based on the specific form of 
$\BPhi(\bx)$ found in (\ref{bphi}) and the 
covariant Stokes theorem that $I+III$ have vanishing 
contribution to the time evolution of $\BPsi$ 
(see 
\cite{static18} for details). 
Then the effective equation for 
$\der_t\Psi_\Sigma$ in (\ref{dtps}) reads 
\begin{align}\label{231}
\ri\der_t \Psi_\Sigma = \gamma_0  \left( -\frac\ka2\der_{\phi\phi} 
+ \ri\gamma^i \der_{i} \phi (\bx) \der_\phi 
+ \frac1\ka V(\phi)) \right) \Psi_\Sigma  
- 
    \ri [\omega_0{}, \Psi_\Sigma ]
= : \hat{H}{}_0 -  
     \ri   [\omega_0{}, \Psi_\Sigma ] . 
\end{align} 
By introducing $\Psi'_\Sigma : = U^{-1} \Psi_\Sigma U$, 
$\; \hat{H}_0':= U^{-1} \hat{H}_0 U$, 
where 
\begin{align}\label{e36}
\begin{split}
U(\bx,t)& = \mathcal{T}e^{ -  \int_0^t \rd s\, \omega_0(\bx,s) } 
\end{split}
\end{align}
is the tranformation determined by the time-ordered exponential, 
we obtain 
\beq \label{52}
\ri\der_t \Psi'_\Sigma = U^{-1} \hat{H}_0 \Psi_\Sigma U 
= \hat{H}_0' \Psi'  
= \gamma'_0  \left( -\frac\ka2\der_{\phi\phi} 
+ \ri\gamma'{}^i \der_{i} \phi (\bx) \der_\phi 
+ \frac1\ka V(\phi) \right) \Psi'~,
\eeq 
where the transformation $\gamma{}^\mu \rightarrow \gamma'{}^\mu$: 
 $\gamma'{}^\mu (x) := U^{-1}(x) \gamma^\mu (x) U (x)$
is an automorphism of the Clifford algebra as  
$\gamma'{}^\mu\gamma'{}^\nu + \gamma'{}^\nu \gamma'{}^\mu = 2 U^{-1}g^{\mu\nu}U 
= 2 g^{\mu\nu} $. 
Using~(\ref{52}) we write 
\begin{align}
\ri \der_t \BPsi &= \Tr \int\! \rd\bx\ 
\left\{ \frac{\delta \BPsi}{\delta \Psi'{}^T_\Sigma (\bx)}  \ri \der_t \Psi'_\Sigma \right\} 
 = \Tr \int\! \rd\bx\, \left\{ \frac{\delta \BPsi}{\delta \Psi'{}^T_\Sigma(\bx)} \hat{H}'_0 \Psi'_\Sigma \right\} .
\end{align}
By comparing it with (\ref{dtps})  and (\ref{eq6n}) we conclude that 
the term $IV$    
is taken into account if the quantities in the expression of 
the wave functional in (\ref{bpsi3}) are replaced by the primed 
($U$-transformed) ones, i.e., 
\beq \label{bpsipr}
\BPsi \sim  
 \Tr
\left \{\BXi{}'([\Psi'_\Sigma];\check{\bx})~
e^{-\ri \phi(\bx)\gamma{}'{}^i\der_i\phi(\bx)/\ka}~
 \frac{\gamma{}'_0}{\sqrt{g} \ka}
 \Psi{}'_\Sigma (\bx) \right \}_{\mbox{\Large $\rvert$} \scriptstyle
 \ka \stackrel{\mbox{\tiny $$}}{\mbox{\scriptsize $\longmapsto$}} 
 \gamma{}'_0\delta(\mathbf{0})/ \sqrt{g}  } . 
\eeq
The fact that this expression is valid for any choice of the point 
$\bx$ allows us to fix $\BXi{}'([\Psi'_\Sigma];\check{\bx})$ and obtain the expression of $\BPsi$ as a formal continual product 
\beq\label{schrod}
\BPsi \sim 
  \Tr \bigg \{\prod_\bx
e^{-\ri\phi(\bx)\gamma'{}^i\der_i\phi(\bx)/\ka} 
 \underline{\gamma}{}_0
 \Psi_\Sigma (\phi(\bx), \bx, t)
\bigg\}{}_{\mbox{\Large $\rvert$} \scriptstyle
 \ka\, 
 \mapsto 
 \gamma'_0\delta(\mathbf{0})/\! \sqrt{g}  }~. 
\eeq 
This symbolic formula expresses the Schr\"odinger wave functional in terms of precanonical wave function restricted to a field configuration $\Sigma$ in general space-time (in the 
adapted coordinates with $g_{0i}=0$). 

The formal continual product expression in~(\ref{schrod}) 
 can be understood as the multidimensional product integral~\cite{prodint,slavik} 
with the invariant measure $\sqrt{h} d\bx$ 
\beq \label{pint}
\BPsi \sim 
  \Tr \left \{   
  \underset{\!\bx}{\scalebox{1.5}{$\displaystyle \prodi$}} 
  e^{-\ri\phi(\bx)\gamma'{}^i (\bx) \der_i\phi(\bx)/\ka}
\underline{\gamma}{}'_0 \Psi'_\Sigma (\phi(\bx), \bx, t)
\right \}{}_{\mbox{\Large $\rvert$} \scriptstyle
 \frac1\ka 
 \mapsto \underline{\gamma}{}'_0
 \sqrt{h}\rd\bx  } , 
\eeq  
where the notation of the product integral of matrix-valued functions $F(\bx)$ 
is used  
\begin{align}
  \underset{\!\bx}{\scalebox{1.5}{$\displaystyle \prodi$}} 
  e^{F (\bx)\rd\bx} 
  = 
  \underset{\!\bx}{\scalebox{1.5}{$\displaystyle \prodi$}} 
  \big(1 +  F (\bx) \rd\bx \big) . 
\end{align}
The expression (\ref{pint}) generalizes our previous result 
in flat space-time \cite{atmp1} to curved space-times.  
In curved space-times, 
the spatial integration measure $\rd \bx$ is replaced by the invariant one $\sqrt{h}\rd\bx$ and the Dirac matrices 
are $\bx$-dependent and nonlocally $t$-dependent if 
$\omega_0\neq 0$.

Note that one-dimensional product integral coincides with the 
well known path- or time-ordered exponential. The miltidimensional 
generalization is more problematic \cite{prodint,slavik}. 
However, the trace of the multidimensional product integral 
can be understood as the continuum limit of the averaging 
of the product of matrices in the infinitesimal cells of space 
over all possible permutations of them in the product over all 
the cells, if the corresponding limit exists:  
\beq \label{lim41}
\Tr\ \underset{\!\bx \in V}{\scalebox{1.5}{$\displaystyle \prodi$}} 
  e^{F (\bx)\rd\bx} 
  := \lim_{N\rightarrow\infty} \frac{1}{N!}\, \Tr 
  \sum_{P(N)} e^{F (\bx_1) \Delta\bx_1} 
  e^{F (\bx_2) \Delta\bx_2} ... e^{F (\bx_N) \Delta\bx_N}~, 
\eeq
where $P(N)$ is the set of all permutations of $(1,2,...,N)$, 
the volume of integration $V\ni \bx$ is partitioned into $N$ small sub-volumes  
$\Delta\bx_1, ... ,\Delta\bx_N$ whose volumes are vanishing 
in the $N\rightarrow\infty$ limit, 
and $F (\bx_i)$ denotes the matrix $F$ at a point 
$\bx_i \in \Delta\bx_i$.


In static space-times when $\omega_0=0$, there is no non-local 
time dependence of the quantities in the expression of 
$\BPsi$ in terms of precanonical wave function and 
equation (\ref{231}) 
can be solved by the Ansatz 
\beq\label{e40}
\Psi_\Sigma = 
e^{+\frac{\ri}{\ka} \phi(\bx)\gamma^i \der_i\phi(\bx)} 
\Phi_\Sigma , 
\eeq  
where $\Phi_\Sigma$ obeys 
\beq \label{e44}
\ri\der_t \Phi_\Sigma = \gammat_0 (\bx) 
\left( -\frac\ka2\der_{\phi\phi} - \frac{1}{2\ka}g^{ij} (\bx) \der_i \phi (\bx) \der_j \phi (\bx) + \frac1\ka V(\phi) \right)  \Phi_\Sigma , 
\eeq
with 
\beq \label{eq42}
\gammat^\mu (\bx):=
e^{-\frac{\ri}{\ka} \phi(\bx)\gamma^i \der_i\phi(\bx)} 
\gamma^\mu (\bx) e^{+\frac{\ri}{\ka} \phi(\bx)\gamma^i \der_i\phi(\bx)}  
\eeq 
and $\gamma^\mu \rightarrow \gammat^\mu $ being a 
local Clifford algebra~automorphism.
From (\ref{e44}) it follows that $\Phi_\Sigma$ can be taken 
in the form $\Phi_\Sigma = (1+\underline{\gamma}{}^0) \Phi^{\times}_\Sigma~
$
where $\Phi^{\times}_\Sigma$ is a scalar function such that 
\beq \label{e101}
\ri \der_t \Phi^{\times}_\Sigma = \sqrt{g_{00}}\left( -\frac\ka2\der_{\phi\phi} - \frac{1}{2\ka}g^{ij} (\bx) \der_i \phi (\bx) \der_j \phi (\bx) + \frac1\ka V(\phi) \right)  
\Phi^{\times}_\Sigma . 
\eeq
In terms of the scalar function $\Phi^\times_\Sigma$ 
 the wave functional (\ref{pint}) reduces to the 
 curvilinear product integral of scalar functions 
\beq
\BPsi \sim 
  \underset{\!\bx}{\scalebox{1.5}{$\displaystyle \prodi$}} \Phi^\times_\Sigma (\phi(\bx), \bx, t) 
 {}_{\mbox{\Large $\rvert$} \scriptstyle
 \frac1\ka 
 \mapsto 
 \sqrt{h}\rd\bx  } , 
\eeq  
which can be defined without any complications related to the definition of the multidimensional product integral of non-commutative matrix functions. 

\newcommand{\defprodint}{ 
OLD TEXT DEFINING TR PRODINT, RMVD 


  Equation~(\ref{fs1}) 
coincides with the canonical functional derivative Schr\"odinger equation~(\ref{fs}). 
Thus the latter is derived from the precanonical Schr\"odinger equation 
 as the limiting 
case corresponding to~(\ref{lim}). In this case,  we can also specify the functional 
$\BXi([\Psi_\Sigma(\bx)], \check{\bx})$ in~(\ref{bpsi3}) by combining the observations 
presented above together and noticing that the relation~(\ref{bpsi3}) is valid at 
{\it any\/} given point ${\bx}$.
This 
is possible only if the functional $\BPsi$ 
 is the continual  product of identical 
 terms at all points ${\bx}$, namely,   
\beq 
\BPsi \sim 
  \Tr \bigg \{\prod_\bx
e^{-i\phi(\bx)\gamma^i\der_i\phi(\bx)/\ka} 
 \underline{\gamma}{}_0
 \Psi_\Sigma (\phi(\bx), \bx, t)
\bigg\}{}_{\mbox{\Large $\rvert$} \scriptstyle
 \ka\, 
 \mapsto 
 \gamma_0\delta(\mathbf{0})/\! \sqrt{g}  }~, 
\eeq 
where $\sim$ means an equality up to a normalization 
factor which includes $\ka$ and $\sqrt{h}$. 

The formal continual product expression in~(\ref{schrod}) 
 can be understood as the multidimensional product integral~\cite{prodint,slavik} 
\beq 
\BPsi \sim 
  \Tr \left \{   
  \underset{\!\bx}{\scalebox{1.5}{$\displaystyle \prodi$}} 
  e^{-i\phi(\bx)\gamma^i (\bx) \der_i\phi(\bx)/\ka}
\underline{\gamma}{}_0 \Psi_\Sigma (\phi(\bx), \bx, t)
\right \}{}_{\mbox{\Large $\rvert$} \scriptstyle
 \frac1\ka 
 \mapsto \underline{\gamma}{}_0
 \sqrt{h}\rd\bx  } , 
\eeq  
where the notation of the product integral of matrix-valued functions $F(\bx)$ 
as proposed by R. Gill~\cite{gill} 
(and implemented in the \LaTeX\ package {\tt prodint}) 
is used  
\begin{align}
  \underset{\!\bx}{\scalebox{1.5}{$\displaystyle \prodi$}} 
  e^{F (\bx)\rd\bx} 
  = 
  \underset{\!\bx}{\scalebox{1.5}{$\displaystyle \prodi$}} 
  \big(1 +  F (\bx) \rd\bx \big) . 
\end{align}
The expression in~(\ref{pint}) generalizes a similar result obtained in flat 
space-time earlier~\cite{atmp2}.  
The only difference is that in curved space-time 
the spatial integration 
measure $\rd \bx$ is replaced by the invariant one $\sqrt{h}\rd\bx$ 
and the Dirac matrices in static space-times 
are $\bx$-dependent. 

In $(1+1)$-dimensional space-time, 
the product integral above is given by the well known path-ordered exponential,  
or the Peano--Baker series 
(also known as the Dyson series in the context of perturbative QFT 
and the path-ordered phase related to the Wilson loop in gauge theory), cf.~Equation~(\ref{e36}) below.   
A multidimensional generalization is briefly discussed in the books~\cite{prodint, slavik} and probably needs further refinement. 
However, in our case, instead of defining the product integral of arbitrary non-commutative matrices, we need only the trace of the product integral of Clifford-algebra valued functions. 
This significantly simplifies the task of defining the expression~(\ref{pint}) mathematically.  
For example, in the one-dimensional case, 
the taking of the trace of each of the terms in the series expansion of the ordered exponential
in~(\ref{e36})) implies 
that the matrices under the integrals in the series expansion 
of the trace of product integral 
are multiplied in the cycling permuted way, which 
can be generalized to the multidimensional case,  rather than a time-ordered one, which implies 
a one-parameter ordering whose multidimensional generalization is problematic. 
Then, if the corresponding limit exists, 
\beq 
\Tr\ \underset{\!\bx \in V}{\scalebox{1.5}{$\displaystyle \prodi$}} 
  e^{F (\bx)\rd\bx} 
  := \lim_{N\rightarrow\infty} \frac{1}{N!}\, \Tr 
  \sum_{P(N)} e^{F (\bx_1) \Delta\bx_1} 
  e^{F (\bx_2) \Delta\bx_2} ... e^{F (\bx_N) \Delta\bx_N}~, 
\eeq
where $P(N)$ denotes all permutations of $(1,2,...,N)$, 
the volume of integration $V\ni \bx$ is partitioned into $N$ small sub-volumes  
$\Delta\bx_1, ... ,\Delta\bx_N$ whose volumes are taken to zero as $N\rightarrow\infty$, 
and $F (\bx_i)$ denotes the matrix $F$ at a point 
$\bx_i \in \Delta\bx_i$. The existence of the limit in~(\ref{lim41}) 
and its independence on the partitioning of $VI$ into $N\rightarrow\infty$ 
sub-volumes $\Delta\bx_i$ and the choice of points  
$\bx_i$ within the subvolumes $\Delta\bx_i$
imply a certain continuity of the dependence of the matrix 
elements of $F$ of $\bx$, 
similarly to the definition of the Riemann integral of functions. 

 } 

\section{\large Conclusion} 

It is demonstrated that in the case of scalar field theory in curved space-time the precanonical Schr\"odinger equation (\ref{crv-ns}) leads to the canonical functional derivative Schr\"odinger equation (\ref{fs}) in the limiting case when $\underline{\gamma}{}_0\ka$  is replaced by $\delta^{\mathrm{inv}}{\bf(0)}$ whose regularized value is the ultraviolet cutoff of the volume of momentum space, whose introduction is one of the ways of defining the second variational derivative at equal points in the canonical
functional derivative Schr\"odinger equation (\ref{fs}).  As a by-product, we also obtain 
the expression of the Schr\"odinger wave functional as the continual product 
or product integral of precanonical wave function restricted to a field 
configuration. In space-times with vanishing zero-th component of spin connection, 
this expression reduces to the product integral of a scalar function obtained from precanonical wave function. In nonstatic space-times with non-vanishing zero-th component of spin connection, the Schr\"odinger  wave functional is expressed in terms of the product integral of a nonlocal transformation of precanonical wave function defined by a time-ordered exponential of the zero-th component of spin connection. 

We hope that these results can stimulate new approaches to the treatment of effects of quantum fields in curved space-time \cite{parker}, such as the Hawking radiation, and they also can lead to better understanding of the nature of states of quantum fields in arbitrary space-times when a separation to positive- and negative-frequency modes is not possible.  They also may help 
to clarify the connections between the existing approaches to quantum gravity originating from the canonical quantization of general relativity \cite{rovelli,kiefer} and precanonical quantization of gravity 
 (see \cite{pqgr,pqgr4,pqgr3,pqgr2,qg,qg2,qg3,qg4}). 

\newcommand{\oldconcls}{ TO IGNORE  
The result of this  paper generalizes to arbitrary space-times 
(with $g_{0i} = 0$) 
the statement of our previous papers~\cite{atmp1,atmp2,romp2018} that the standard functional Schr\"odinger representation of QFT is a certain 
 singular limiting case of the theory of quantum fields obtained by precanonical quantization.

The symbolic or singular nature of the limiting transition from precanonical quantization 
to the standard formulation of QFT in functional Schr\"odinger representation 
is related to the fact that the latter, due to the presence of the second functional  
derivative at equal points, is not a well-defined theory unless a regularization is introduced. 
The regularization typically introduces a UV cutoff scale as an additional element of the theory 
removed by a subsequent renormalization. In precanonical quantization, 
the  ultraviolet scale $\varkappa$ appears as an inherent element quantization, 
which, unlike other theories introducing an ultraviolet fundamental length or cutoff, 
does not alter the relativistic space-time at smaller scales. 

This rises a question: is $\varkappa$ a fundamental scale or an auxiliary element of precanonical quantization of fields.  
On the one hand,  in free scalar theory,  $\varkappa$ disappears from the observable characteristics of a quantum field because the spectrum of DW Hamiltonian operator is proportional to $\varkappa$.   However,  in interacting scalar theory, powers of $\varkappa$ enter in the perturbative corrections to the spectrum of DW Hamiltonian $\hat{H}$ thus suggesting that $\varkappa$  can be renormalized away by  absorbing the expressions  with the bare mass, bare coupling constant  and $\varkappa$  in the "observed mass". On the other hand, an estimation of the mass gap in  $(3+1)$-dimensional quantum pure SU(2) gauge theory derived by precanonical quantization: 
$\Delta m \gtrsim 0.86 (g^2\ka)^{1/3}$ {}~\cite{ym-mass} ($g$ is the bare gauge coupling constant)   and a naive estimation of the cosmological constant based on the precanonically quantized pure Einstein gravity~\cite{qg} seem to consistently point to the  very rough estimation of the scale of $\varkappa$ at $\sim 10^2 MeV$, 
i.e. well below the Planck scale.   We hope to clarify this surprising fact and the nature of $\varkappa$ in our future work.  
 } 


\end{document}